\documentclass{article} 
\usepackage{iclr2023_conference,times}

\usepackage{amsmath,amsfonts,bm}

\def\eqref#1{equation~\ref{#1}}

\def\1{\bm{1}}

\DeclareMathAlphabet{\mathsfit}{\encodingdefault}{\sfdefault}{m}{sl}
\SetMathAlphabet{\mathsfit}{bold}{\encodingdefault}{\sfdefault}{bx}{n}

\usepackage{parallel,enumitem}
\usepackage{hyperref}
\usepackage{url}
\usepackage{paralist}
\usepackage{times}
\usepackage{latexsym}
\usepackage{graphicx}
\usepackage{color,soul}
\graphicspath{{./images/}}
\usepackage{booktabs}
\usepackage{tabularx}
\usepackage{caption}
\usepackage{subcaption}
\usepackage[export]{adjustbox}
\usepackage{arydshln}
\usepackage{wrapfig,lipsum}
\usepackage{adjustbox}

\usepackage{listings}
\usepackage{amsmath}

 \usepackage[frozencache=true,cachedir=minted-cache]{minted}
\title{TASTy: A \underline{T}ransformer based \underline{A}pproach to \underline{S}pace and \underline{T}ime complexit\underline{y}}

\author{Kaushik Moudgalya \\ 
University of Montreal\\
Montreal, Quebec, Canada \\
\And
Ankit Ramakrishnan \\
Northeastern University \\
Boston, Massachusetts, USA \\
\And
Vamsikrishna  Chemudupati\\
University of Montreal\\
Montreal, Quebec, Canada \\
\And
Xing Han Lu \\
McGill University \\
Montreal, Quebec, Canada \\
}

\iclrfinalcopy 
\begin{document}

\maketitle

\begin{abstract}
Code based Language Models (LMs) have shown very promising results in the field of software engineering with applications such as code refinement, code completion and generation.
However, the task of time and space complexity classification from code has not been extensively explored due to a lack of datasets, with prior endeavors being limited to Java. In this project, we aim to address these gaps by creating a labelled dataset of code snippets spanning multiple languages (Python and C++ datasets currently, with C, C\#, and JavaScript datasets being released shortly).  We find that existing time complexity calculation libraries and tools only apply to a limited number of use-cases. The lack of a well-defined rule based system motivates the application of several recently proposed code-based LMs. We demonstrate the effectiveness of dead code elimination and increasing the maximum sequence length of LMs. In addition to time complexity, we propose to use LMs to find space complexities from code, and to the best of our knowledge, this is the first attempt to do so. Furthermore, we introduce a novel code comprehension task, called cross-language transfer, where we fine-tune the LM on one language and run inference on another. Finally, we visualize the activation of the attention fed classification head of our LMs using Non-negative Matrix Factorization (NMF) to interpret our results.
\end{abstract}

\section{Introduction}
Language Models such as BERT, GPT-3, and T5 \citep{BERT, GPT-3, t5} have propagated the idea of pre-training and finetuning due to their success on a variety of downstream tasks. This has led to a plethora of pre-trained models for programming languages (PL). These are used on code related downstream tasks such as code search \citep{cptcodeM}, code translation \citep{CodeT5}, code summarization \citep{code-summarization} etc. We propose to work on one such downstream task: predicting time and space complexities based on code.

Empirical methods of predicting time complexity involve running the same piece of code over differently sized inputs and measuring the time taken for execution, which enables us to approximate a function. Our attempt at classification will help reduce the time and effort required in this process. Additionally, we hope our attempt serves as a stepping stone for future efforts that will remedy this issue once and for all. Finally, we wish to put code-based LM's code comprehension to the test by testing its cross-language transfer capabilities, where we train our model to predict time complexities on one language and evaluate on another.  For even an average human programmer, finding the time complexity for an arbitrary piece of code can be a cumbersome task. Our long-term goals are two-fold:
\begin{inparaenum}
  \item Utilization of LM based complexity predictors on competitive coding platforms, reducing the number of code runs over differently sized inputs.
  \item Integration of methods like ours with AI programming assistants, which will make it easier for programmers to optimize their code and ease the code review process.
\end{inparaenum}

\begin{wrapfigure}{l}{0.5\textwidth}
  \centering
  \includegraphics[trim={0 0 13cm 0},clip,width=0.5\textwidth]{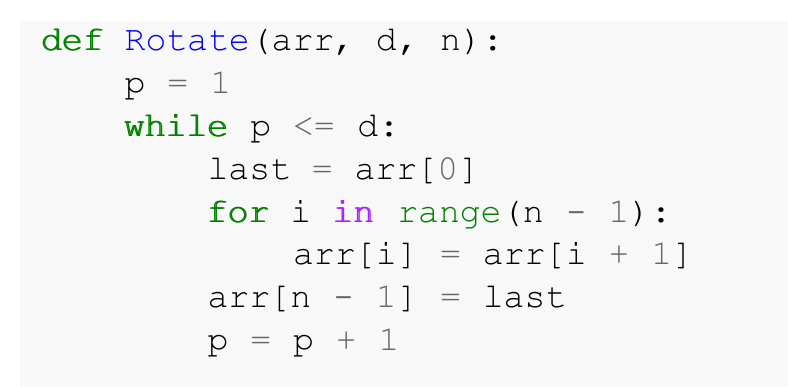}
  \caption{Sample input program for time and space complexity calculation. The expected time and space complexity model outputs would be \textbf{Quadratic} and \textbf{Constant} respectively.}
  \label{sample-code1}
\end{wrapfigure}

Previous datasets with code and its time complexities, such as CoRCoD \citep{Jagriti} and CodeComplex \citep{jeon2023deep} focus solely on Java. In this paper, we introduce a C++ and Python dataset of code with both their corresponding time and space complexities. This is the first dataset of its kind for these languages with time complexities and as far as we know, the first dataset ever to contain space complexities.  A sample input and output of our model are shown in \hyperref[sample-code1]{Figure 1}.\footnote{We plan to release similar datasets for C, C\# and JavaScript shortly.}

Most prior attempts to find time complexities are based on traditional machine learning based methods such as Random Forests, Support Vector Machines \citep{Jagriti} and Light Gradient Boosting Machines \citep{seifa} bolstered by manually constructed features. We believe learned attention weights might fare better since the LMs can benefit from being exposed different code based contexts which is infeasible with constructed features. Hence we evaluate several code-based LMs on our dataset as well as the CodeComplex dataset.

The sole other attempt using LMs for time complexity prediction- \citet{jeon2023deep} uses techniques such as hierarchical attention and dead code elimination. Often, the length of the code exceeds the maximum sequence length of a transformer, which means that it cannot attend to all the tokens in the code. Dead code elimination attempts to solve this problem by removing unused methods and variables, which reduces the length of the code enabling transformers to attend to more tokens. While verifying the efficacy of such elimination, we wish to determine whether increasing sequence lengths \citep{longformer} increases the accuracy of our models.

If LMs can truly understand code at a conceptual level, models pre-trained on multiple languages should be able to predict time complexity when we fine-tune them on say, our Python based dataset and evaluate them on our C++ based dataset. We believe this is reasonable since a competent human programmer who has experience in multiple programming languages, would be able to calculate time complexities for programs in all those languages. 

Finally, we use Non-negative Matrix Factorization (NMF) to decompose the activations of the Feed Forward Neural Network (FFNN) to extract interpretable components as described in \citet{alammar-2021-ecco}. NMF affords researchers to probe parts-to-whole relationships as described in \citet{Lee1999}. We use this quality of NMF to qualitatively analyse the trained models. 


We also present a survey of existing tools that predict time and space complexity and detail our findings in \hyperref[extant-complexity-tools]{Appendix A.1}. Unfortunately, none of the tools we found provide a comprehensive solution to the problem of identifying algorithm complexities.

\section{Related Work}
\label{sec:Related Works}

\textbf{CoRCoD}: 
\citet{Jagriti} created a dataset of 933 Java codes and their complexities. We extend their work by making a dataset of C++ and Python programs. Like our dataset, their code also seems to stem from GeeksforGeeks based on the comments in the code. They classify programs into 5 classes: $O(1)$, $O(n)$,  $O(\log n)$, $O(n \log n)$, $O(n^2)$. Their methods include engineered features such as number of loops, number of if statements, maximum depth of nested loops etc. extracted from the Abstract Syntax Tree. Additionally, they also use graph2vec which uses a skipgram \citep{word2vec} model to compute graph embeddings corresponding to code embeddings, which are then classified using traditional ML methods such as Random Forests, Support Vector Machines etc. 

\textbf{\citet{seifa}}: Uses IBM’s CodeNet Dataset to predict time complexities. They derive the labels by defining programs as sub-polynomial, polynomial or above-polynomial (dataset is not publicly available). Like CorCoD, they engineer features such as number of loops, breaks, if statements along with graph based representations derived from an Abstract Syntax Tree. Their best performing approach uses a 15 layer residual network which works on manually constructed features concatenated with graph representations.

\textbf{CodeComplex}: \citep{jeon2023deep} construct a human annotated Java dataset (\href{https://huggingface.co/datasets/codeparrot/codecomplex}{CodeComplex}) of 3803 Java codes and their time complexities (our dataset is comprised of C++ and Python). We also note that this dataset has some overlap with CoRCoD. Their approach combines dead-code elimination, multi-level pre-training objectives such as loop depth prediction, number of parameters prediction along with hierarchical attention which generates method level and class level embeddings which are combined to create a final embedding for prediction. 

\section{Our Dataset}
We call our dataset the GeeksforGeeks (GFG) \footnote{\href{https://www.geeksforgeeks.org/}{https://www.geeksforgeeks.org/}} dataset, named after the site where the code was scraped. Our C++ and Python dataset contain 1410 and 1373 codes and their corresponding time and space complexities respectively. Check \hyperref[dataset-comparison-table]{Table 1} for dataset class distribution. We remark that our dataset is imbalanced and the linear class  alone constitutes 54.6\% and 55.8\% of our Python and C++ time complexity datasets, while combining the linear and quadratic classes constitutes 82.3\% of both datasets. There is a tendency on coding interview preparation platforms such as GFG to curate more solutions around these complexities in order to be more instructive. We do not exhaustively scrape the GFG website, thus, we could potentially fix the high class imbalance with selective scraping. Other statistics about the dataset such as construction details, the space complexity class breakdown and statistics about the code length and its distribution are mentioned in the \hyperref[dataset-construction-details]{Appendix A.2} and \hyperref[dataset-additional-stats]{Appendix A.3} respectively.

\begin{table}
\centering
\begin{tabular}{ccccc}
\toprule
\textbf{Time Complexity} & \multicolumn{1}{l}{\textbf{CoRCoD (Java)}} & \multicolumn{1}{l}{\textbf{CodeComplex (Java)}} & \multicolumn{1}{l}{\textbf{GFG (C++)}} & \multicolumn{1}{l}{\textbf{GFG (Python)}}  \\ 
\midrule
$O(1)$      & 143  & 533   & 33     & 35   \\
$O(n)$   & 385    & 472 & 787  & 750 \\
$O(n^2)$  & 200 & 553 & 374  & 381    \\
$O(n^3)$   & -     & 579 & 35    & 36 \\
$O(\ln n)$  & 55   & 576  & 26  & 25  \\
$O(n \ln n) $  & 150   & 518  & 127  & 121 \\
NP-hard  & -    & 572   & 28   & 25\\ 
\toprule
\textbf{Total} & 933     & 3803   & 1410 & 1373 \\
\bottomrule
\end{tabular}
\caption{Comparison of the per class occurrences of time complexities across different datasets CodeComplex and CoRCoD numbers borrowed from respective papers.}
\label{dataset-comparison-table}
\end{table}

\section{Experiments}
We consider BERT, CodeBERT, GraphCodeBERT, CodeT5 \citep{BERT,codebert,graphcodebert, CodeT5} and Longformer \citep{longformer} as our candidate models. CodeBERT, GraphCodeBERT and CodeT5 have been pre-trained on  CodeSearchNet \citep{codesearchnet}, a dataset comprising of 6 languages: Python, Java, JavaScript, PHP, Ruby and Go. CodeT5 has additionally been trained on C and C\# which their work sources from BigQuery \footnote{\href{https://console.cloud.google.com/
marketplace/details/github/github-repos}{https://console.cloud.google.com/
marketplace/details/github/github-repos}}. All the models are fine-tuned by adding a linear layer (a classification head) at the end to obtain the desired output. In addition to fine-tuning these models on our GFG Python and GFG C++ datasets, we also fine-tune these models for time complexity classification on the CodeComplex dataset. 

For fine-tuning, we utilize the Adam \citep{Adam} optimizer with an effective batch-size of 32; a default batch-size of 8 and 4 steps of gradient accumulation \citep{gradient-accumulation}. Across all runs, we fine-tune for 15 epochs using a constant 1e-5 learning rate. We also employed mixed-precision training for all models \citep{mixed-precision-training} and gradient checkpointing \citep{gradient-checkpointing} for all models except CodeBERT. We use the base version of all the models. All of our results are averaged across 5 runs.

For the CodeComplex dataset, we use the training and testing splits provided by \citet{jeon2023deep} with dead code elimination. For our GFG dataset, we use a 80-20 label stratified train-test split for both time and space. 

We consider the BERT results to be our baseline, since it is trained only on Natural Language while the other models mentioned are trained either on both Natural and Programming Languages or only Programming Languages. While the Longformer too has only been trained on Natural Language, it has an inherent advantage over BERT since it can attend to longer sequence lengths.

\section{Results}
\subsection{Time \& Space complexity results with constant maximum sequence length}
We fine-tune the models mentioned in \hyperref[table2]{Table 2} except the Longformer with a constant maximum sequence length of 512 tokens. We expected CodeT5 to achieve the best accuracy since it has a higher number of learnable parameters, but GraphCodeBERT achieves the best accuracy across all 3 datasets, on both time and space. While the disparity in accuracies between BERT and the programming language based models is quite significant on the CodeComplex dataset, the disparity on the GFG C++ and GFG Python datasets is not so drastic. In fact, CodeT5 does even worse than BERT on the GFG Python dataset, on both time and space. We also report the average per-class accuracies across 5 runs in \hyperref[per-class-accuracy]{Appendix A.4}.

\begin{table}
\centering
\caption{Time and Space Complexity prediction results. We average test accuracies across 5 runs of
all LMs. For Longformers, we vary the max. sequence length. }
\label{table2}
\begin{tabular}{llcccc} 
\toprule
\textbf{\small{Language (Dataset)}} & \textbf{\small{Model}} & \begin{tabular}[c]{@{}c@{}}\textbf{\small{Max. Seq. }}\\\textbf{\small{Length}}\end{tabular} & \textbf{\small{Parameters}} & \begin{tabular}[c]{@{}c@{}}\textbf{\small{Accuracy}}\\\textbf{\small{Time (\%)}}\end{tabular} & \begin{tabular}[c]{@{}c@{}}\textbf{\small{Accuracy}}\\\textbf{\small{Space (\%)}}\end{tabular}  \\ 
\midrule
Java (CodeComplex) & BERT & 512 & 110M & 84.68 & - \\
Java (CodeComplex) & CodeBERT & 512 & 125M & 90.54 & - \\
Java (CodeComplex) & GraphCodeBERT & 512 & 125M & \textbf{92.08} & -\\
Java (CodeComplex) & CodeT5 & 512 & 220M & 88.59 & - \\ 
\hdashline
C++ (GFG) & BERT & 512 & 110M & 75.60 & 74.75 \\
C++ (GFG) & CodeBERT & 512 & 125M & 74.46 & 75.88\\
C++ (GFG) & GraphCodeBERT  & 512 & 125M & \textbf{78.29} & \textbf{79.71} \\
C++ (GFG) & CodeT5  & 512 & 220M  & 71.56   & 70.28 \\ 
\hdashline
Python (GFG) & BERT & 512    & 110M  & 75.92   & 72.58  \\
Python (GFG)  & CodeBERT  & 512   & 125M  & 74.10  & 73.23 \\
Python (GFG)  & GraphCodeBERT  & 512 & 125M  & \textbf{77.89}   & \textbf{73.96}   \\
Python (GFG)   & CodeT5 & 512  & 220M   & 68.43  & 65.16   \\ 
\midrule
Java (CodeComplex)   & Longformer & 256 & 41M     & 80.33  & -       \\
Java (CodeComplex)  & Longformer  & 512   & 41M   & 85.62      & -   \\
Java (CodeComplex) & Longformer  & 1024    & 41M & 85.75      & -   \\
Java (CodeComplex)          & Longformer     & 2048                                                                         & 41M                 & \textbf{86.66}                                                                          & -                                                                               \\ 
\hdashline
C++ (GFG)                   & Longformer     & 256                                                                          & 41M                 & 71.77                                                                          & -                                                                               \\
C++ (GFG)                   & Longformer     & 512                                                                          & 41M                 & \textbf{72.55}                                                                          & -                                                                               \\
\bottomrule
\end{tabular}
\end{table}


\subsection{Time complexity results with varying maximum sequence length}
We fine-tune the Longformer model across a range of maximum sequence lengths as shown in latter part of \hyperref[table2]{Table 2} and \hyperref[fig:longformer]{Figure 2}. On both the GFG C++ and the Java CodeComplex dataset, there is an increase in complexity prediction accuracy with increasing sequence length. 

\noindent 
    \begin{minipage}[t]{0.48\textwidth}
        \adjustimage{width=1\textwidth,height=4cm,valign=t,caption={mycaption}}{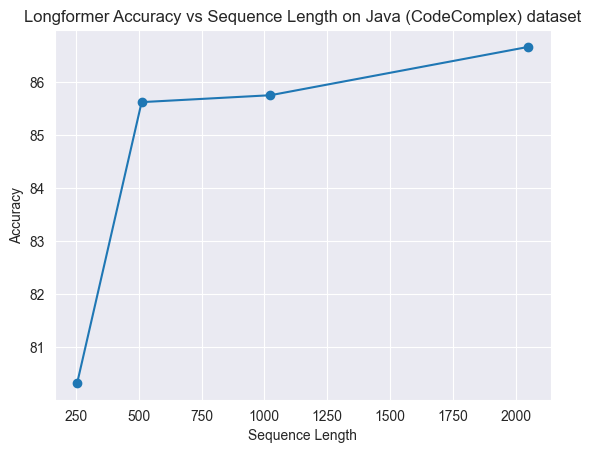}
        \captionof{figure}{Increasing accuracy with increasing maximum sequence length using Longformers on the Java CodeComplex dataset.}
        \label{fig:longformer}
    \end{minipage}
    \hfill 
    \begin{minipage}[t]{0.5\textwidth}
        While there is an observable increase in the accuracy with increasing sequence length on the CodeComplex dataset, the extent of the improvement seems to fizzle out. Similarly, on the GFG C++ dataset, the jump in accuracy while going from 256 to 512 tokens is not comparable to the jump on the CodeComplex dataset. However, note that Longformer is a Natural Language (NL) model, so the fact that the accuracy does increase with increasing sequence length is quite significant. In future work, we'd like to try exploring a PL based model with larger sequence lengths.
    \end{minipage}

    \medskip 

\subsection{Cross language transfer (CLT)}
The results for CLT are given in \hyperref[cross-lang]{Table 3}. BERT has the best accuracies for cross-language transfer on both the target languages, followed by GraphCodeBERT on Python and Longformer (512) on C++.

\begin{table}[h]
\centering
\caption{Cross-language transfer results: Inference of models trained on the CodeComplex (all sources are java) to  target GFG Python (py) / C++ (cpp) datasets. No explicit fine-tuning was performed on the target language.}
\label{cross-lang}
\begin{tabular}{lccc} 
\toprule
\textbf{Model} & \textbf{Parameters} & \begin{tabular}[c]{@{}c@{}}\textbf{Max. }\\\textbf{Seq. Length}\end{tabular} & \begin{tabular}[c]{@{}c@{}}\textbf{Accuracy}\\\textbf{py / cpp (\%)}\end{tabular}  \\ 
\midrule
BERT           & 110M                & 512                                                                          & \textbf{29.13} / \textbf{23.54}                                                    \\
CodeBERT       & 125M                & 512                                                                          & 19.37 / 16.80                                                                      \\
GraphCodeBERT  & 125M                & 512                                                                          & \textbf{23.96} / 18.22                                                                      \\
CodeT5         & 220M                & 512                                                                          & 18.06 / 12.05                                                                      \\ 
\midrule
Longformer     & 41M                 & 256                                                                          & 20.83 / 18.22                                                             \\
Longformer     & 41M                 & 512                                                                          & 15.73 / \textbf{20.92}                                                             \\
Longformer     & 41M                 & 1024                                                                         & 13.83 / 13.19                                                                      \\
Longformer     & 41M                 & 2048                                                                         & 13.98 / 16.87                                                                      \\
\bottomrule
\end{tabular}
\end{table}

\subsection{Dead code elimination}
Introduced by \citet{jeon2023deep}, this technique removes variables, functions etc. that are defined but not used, since scraped code need not be optimal code. We observe a clear increase in the accuracy of LMs with dead code elimination on the CodeComplex dataset \hyperref[dead-code]{Table 4}, thus verifying the results presented by \citet{jeon2023deep}. They report that CodeBERT model's accuracy jumps from 86.0\% to 96.0\% with pre-training, dead code elimination and additional pre-training objectives. We ablated for the accuracy of dead code elimination in our work.

\begin{table}[hbt!]
\centering
\caption{Ablating for the efficacy of Dead Code (DC) elimination on the CodeComplex dataset. Values without DC borrowed from \citet{jeon2023deep}. Values with DC are from \hyperref[table2]{Table 2}.}
\label{dead-code}
\begin{tabular}{lcc} 
\toprule
\multicolumn{1}{c}{\textbf{Model}} & \begin{tabular}[c]{@{}c@{}}\textbf{Accuracy}\\\textbf{w/o DC (\%)}\end{tabular} & \begin{tabular}[c]{@{}c@{}}\textbf{Accuracy }\\\textbf{w/DC (\%)}\end{tabular}  \\ 
\midrule
CodeBERT                           & 86.0                                                                            & 90.54                                                                           \\
GraphCodeBERT                      & 80.3                                                                            & 92.08                                                                           \\
CodeT5                             & 85.9                                                                            & 88.59                                                                           \\
\bottomrule
\end{tabular}
\end{table}

\subsection{Qualitative analysis}
We perform qualitative analyses based on the activations of two models, CodeBERT and GraphCodeBERT. NMF is particularly suited to analysing text qualitatively since it is able to provide us with interpretable components. NMF has found success in medical research \citep{10.1093/bib/bbac246} and information retrieval \cite{10.1145/1816041.1816094}. In our case, we apply NMF to the activations of the neurons across the FFNN layers. i.e. the classification head on top of the LM.

\begin{figure}[h]
\centering
\begin{subfigure}{.5\textwidth}
  \centering
  \includegraphics[width=1.0\columnwidth]{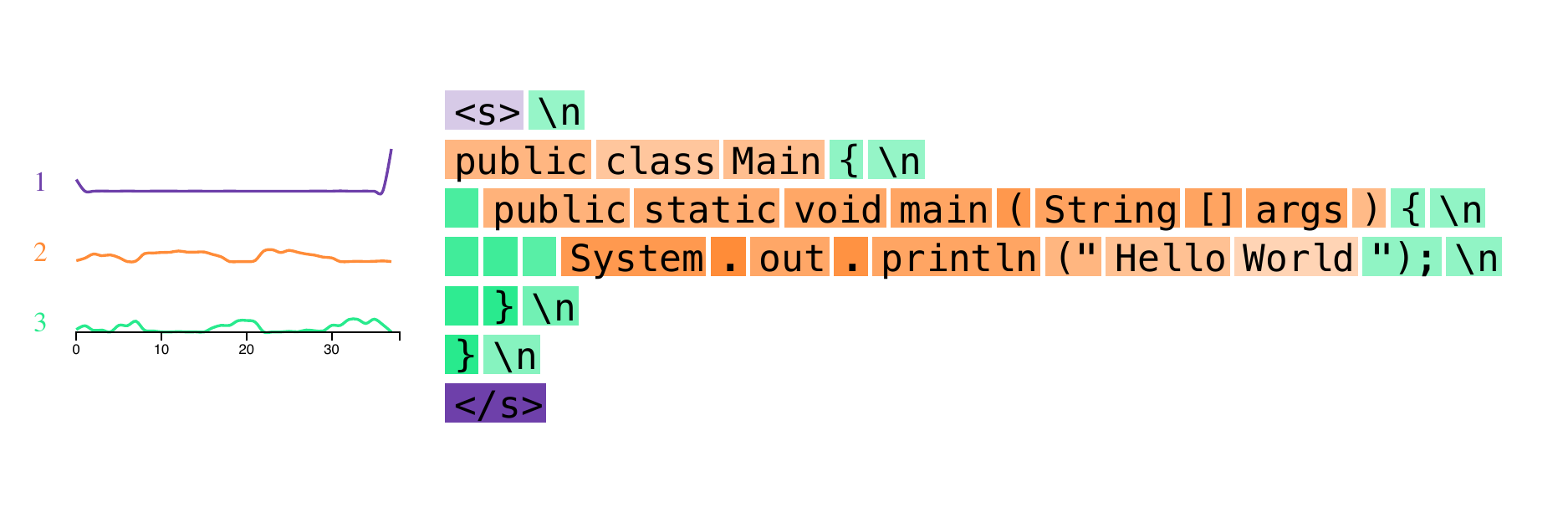}
  \subcaption{3 components}
\end{subfigure}%
\begin{subfigure}{.5\textwidth}
  \centering
  \includegraphics[width=1.0\columnwidth]{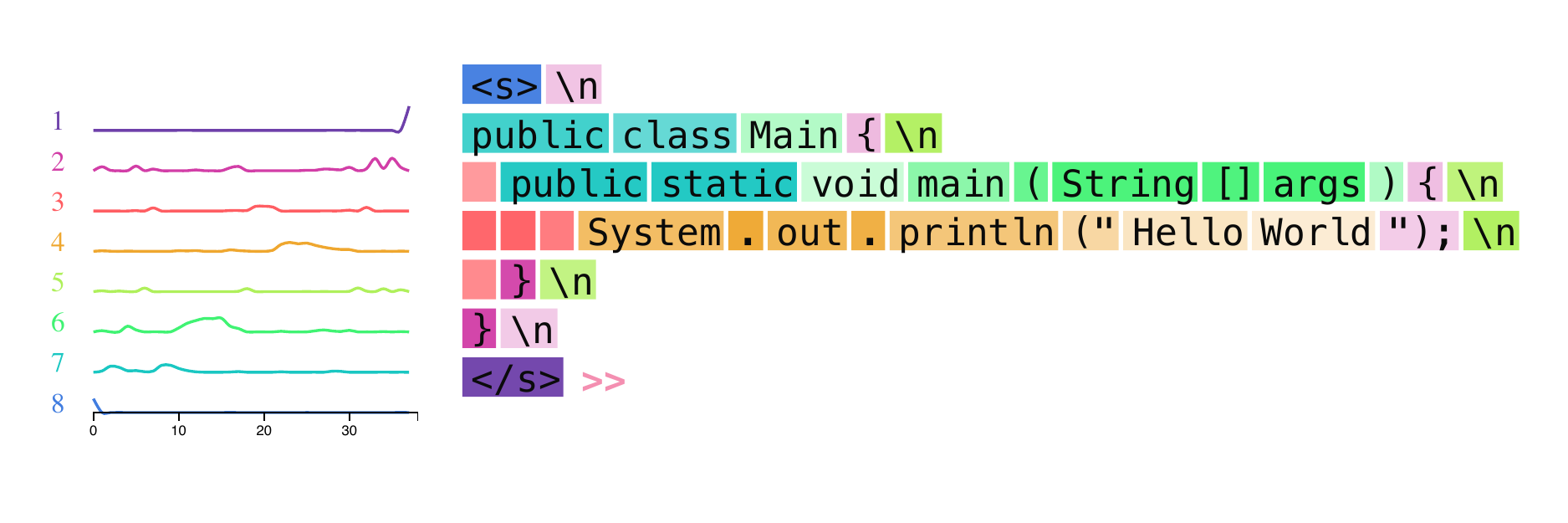}
  \subcaption{8 components}
\end{subfigure}
\caption{Comparing visualizations with different number of components with CodeBERT activations across all layers}
\label{fig:n-components}
\end{figure}

The visualization in \hyperref[fig:n-components]{Figure 3} is an example of using this method \citep{alammar-2021-ecco} on activations produced by CodeBERT after passing a "hello world" java program through it. It is clear by looking at the different components that each component is activated by different items of syntax. For example, component 1 in the figure is closely associated with the start line character \verb"<s>" and the character \verb"</s>". Component 7 is closely linked with access specifiers in the code (\verb"public class" and \verb"public static").

Identifying the number of components one should use for analysis depends on the total length and complexity of the input. NMF solutions are non-unique and we can get dissimilar results across different iterations. Depending on the complexity of the input activations, we vary the number of components to reveal fine-grained or global features of the input text. We compare models across the three dimensions that follow:


\textbf{Number of Components} (\hyperref[fig:n-components]{Figure 3}): Comparing the visualizations produced by a 3 component NMF and an 8 component NMF reveals that different components encode for different kinds of syntax tokens. In the 3 component decomposition, we observe three distinct sections that activate the neurons, the start and end tokens; whitespace tokens such as newlines and tabs; and the main body of the functions. With 8 components we observe an increase in the different semantic units that different components react to. We now see the emergence of components that react to access modifiers, identifier names, parenthesis and individual functions.

\begin{figure}[h!]
\centering
\begin{subfigure}{.5\textwidth}
  \centering
  \includegraphics[width=1.0\columnwidth]{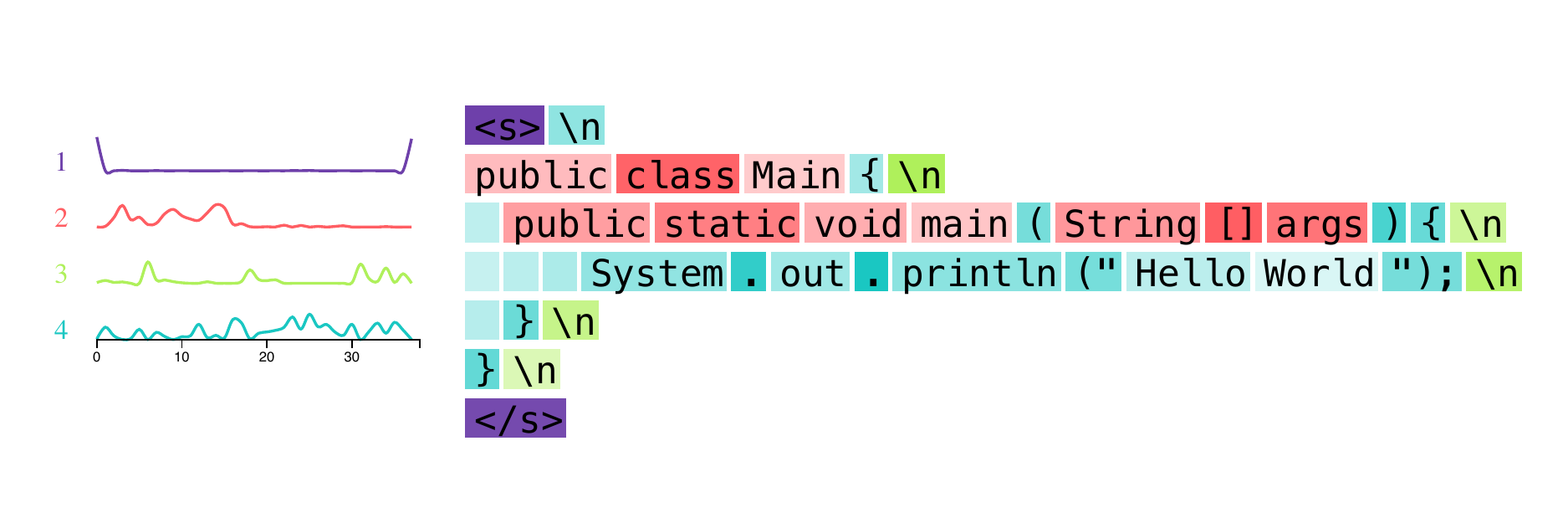}
  \subcaption{Layer 2 to 3 with 4 components}
  \label{fig:test2}
\end{subfigure}%
\begin{subfigure}{.5\textwidth}
  \centering
  \includegraphics[width=1.0\columnwidth]{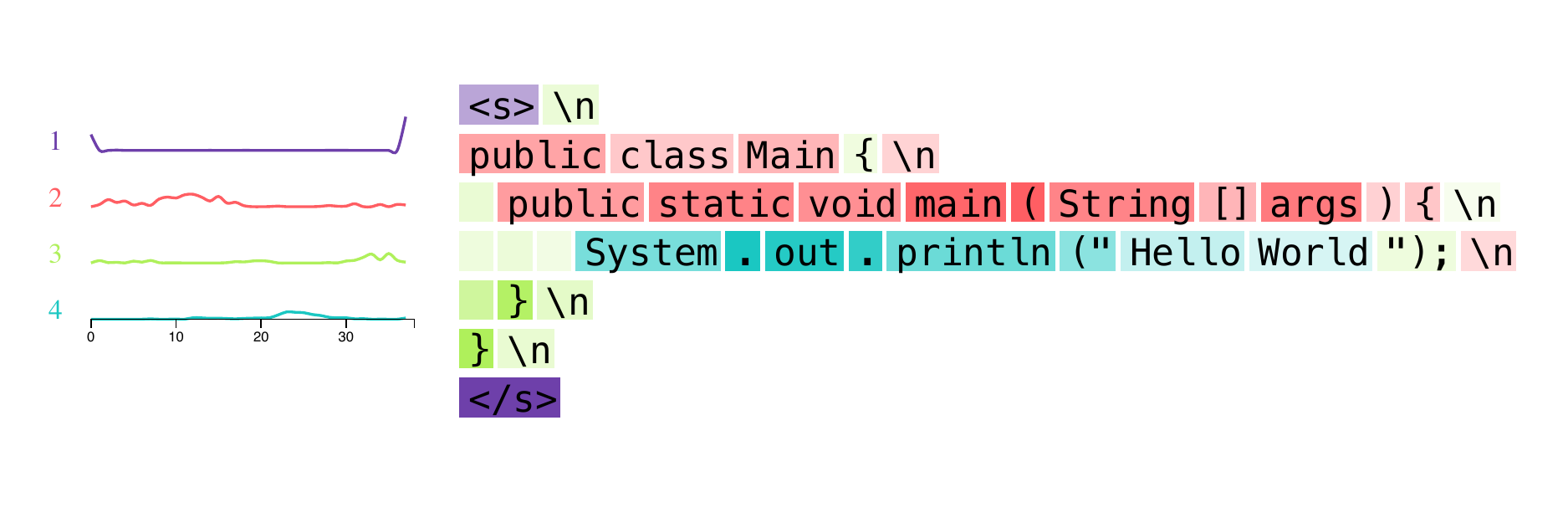}
  \subcaption{Layer 9 to 10 with 4 components}
  \label{fig:test3}
\end{subfigure}
\caption{Comparing visualizations of activations across different layers}
\label{fig:n-layers}
\end{figure}

\textbf{Layers}  (\hyperref[fig:n-layers]{Figure 4}): Activations across the earlier layers result in components with more variation and higher weights on average. Later layers are more smooth and have lower weights. However, the semantic information conveyed by the components appears to be identical.

\begin{figure}[h]
    \centering
      \includegraphics[width=0.70\columnwidth, height=5cm
      ]{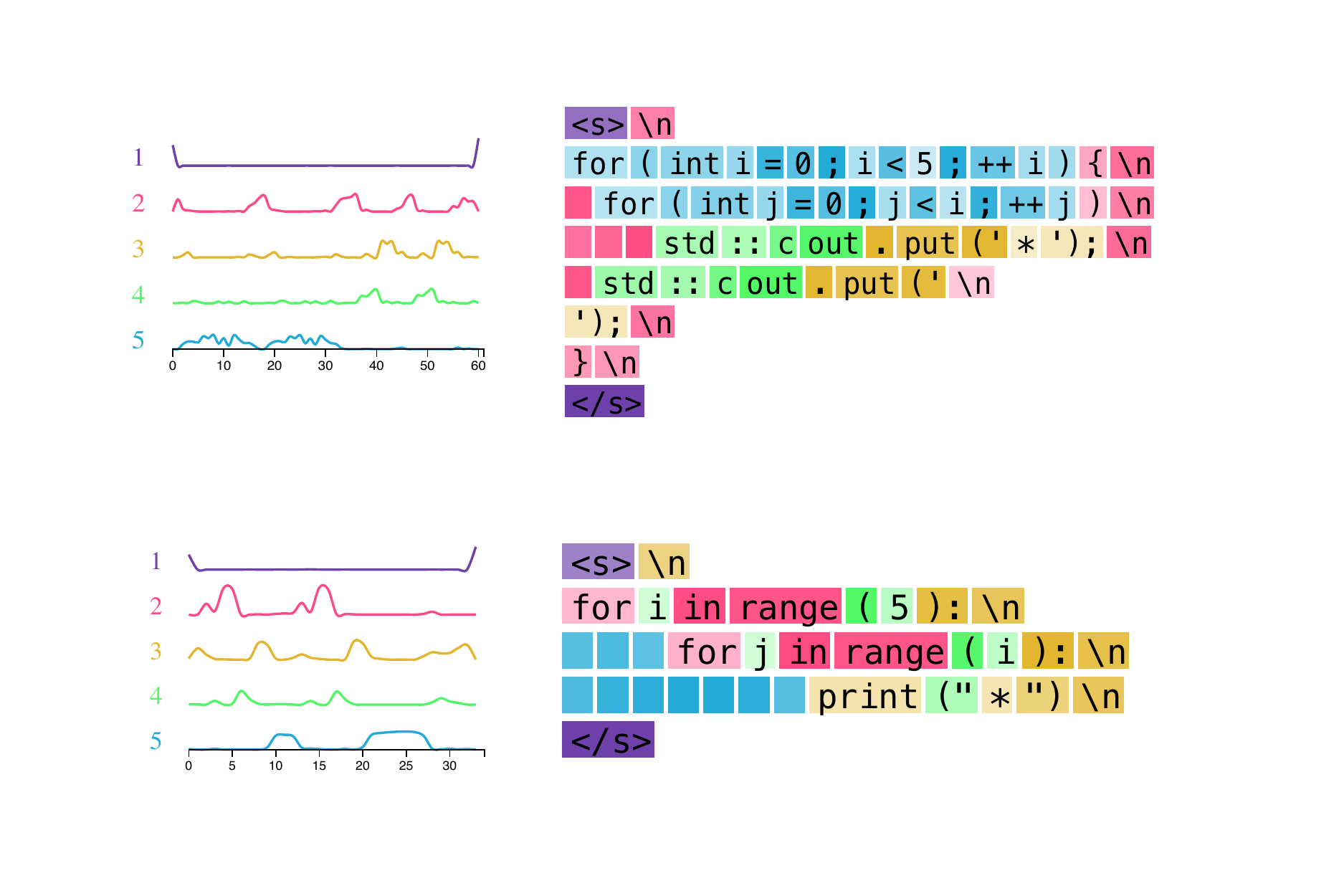}
    \caption{Comparing visualizations of activations where GraphCodeBERT was trained on python. Here we test the model on C++ and Python code}
    \label{fig:n-langs}
\end{figure}

\textbf{Pre-Training Language} (\hyperref[fig:n-langs]{Figure 5}) : We observe visualizations of layer activations for GraphCodeBERT trained on Python. To generate these activations we use a C++ sample and a Python sample. The components for both pairs of languages appear to encode for similar things. For example, component 1 encodes the start and end tokens, components 5 in C++ and 2 in Python both correspond to for-loop structures.

\section{Discussion}
\textbf{Comparison to traditional ML methods.} In contrast to traditional ML methods such as LGBM and Random Forests, we comprehensively explored LMs for our complexity prediction task. We note that traditional ML approaches are not easily parallelizable and cannot be trained on hundreds of thousands of data points concurrently while such an impediment does not exist for LMs \citep{vaswani2017attention}. We also note that ML based approaches require extensive feature engineering such as counting the number of ifs, loops, and breaks, finding nested loop depths etc \citep{Jagriti}. LMs have an inherent scalability benefit along with the benefit of context without the need to create manual features.

\textbf{Natural Language vs. Programming Language models.} While there is a trend of increasing accuracy with increasing sequence length for Longformers on the CodeComplex dataset (\hyperref[table2]{Table 2}, \hyperref[fig:longformer]{Figure 2}), we note that even Longformer (2048 sequence length) cannot surpass the accuracy of GraphCodeBERT (512 sequence length). Observe that all the programming language (PL) based models (CodeBERT, GraphCodeBERT and CodeT5) beat the Natural Language (NL) based models (BERT and Longformer) on the CodeComplex dataset (\hyperref[table2]{Table 2}). PL models also beat NL models on the GFG C++ and Python dataset on both time and space accuracies (\hyperref[table2]{Table 2}) with CodeT5 being the sole exception. We also note the efficacy of dead-code elimination which leads to at least a 2.69\% increase in accuracy across all LMs (\hyperref[dead-code]{Table 4}). 

\textbf{CodeT5.}  Remark that CodeT5 is even outdone by BERT on both the GFG C++ and Python datasets (\hyperref[table2]{Table 2}). Furthermore, despite having a higher number of parameters than all the other models used here, it does worse than even the Longformer having 5x fewer parameters. On an unbalanced dataset, even majority class predictors will achieve high accuracies, hence we will compare CodeT5 on the balanced CodeComplex dataset. We would expect CodeT5 to surpass GraphCodeBERT's performance on CodeComplex since it has additional pre-training objectives, a higher number of parameters and extra training data (only model actually trained on C and C\# which are similar to C++). Instead, we believe that this provides further impetus to the arguments presented by \citet{CodeT5}, where they propose that the Bimodal Dual Generation pre-training objective, while being great for PL-NL tasks such as code summarization and code generation, might bias the model towards such tasks. At its core, time and space complexity prediction are closer to a PL-PL task than a PL-NL task. Hence, it might be interesting to see whether CodeT5 pre-trained without the Bimodal Dual Generation objective continues to do worse than BERT on our dataset. 

\textbf{Python vs. C++ Cross Language Transfer.} We also note that Java to Python CLT accuracies are better than Java to C++ CLT accuracies with two exceptions \hyperref[cross-lang]{Table 3}, the Longformer (512) and Longformer (2048) results. This is explained by the fact that  the pre-training of all code based LMs used included Python data but not C++ data (Code T5 uses C and C\# data additionally). This explains the outliers as well since it stems from a NL based LM (Longformer) rather than a PL based LM, a pattern that is also supported by the BERT results in \hyperref[cross-lang]{Table 3}, since BERT does better than CodeBERT, GraphCodeBERT and CodeT5. We expected that Java to C++ CLT would be easier than Java to Python CLT since syntactically Java as a language is closer to C++ than it is to Python. 

\textbf{NL vs. PL models on Cross Language Transfer.} We conduct a completely novel kind of code-understanding experiment, the CLT experiments (\hyperref[cross-lang]{Table 3}). We expected our NL based models to give us almost random accuracies on this task, which would amount to 14.28\% (1 in 7); i.e. PL based LMs would do better than NL based LMs. However, BERT was the best model on this task across both target languages despite having not being pre-trained on programming languages. Furthermore, the next best performer on Java to C++ CLT was the Longformer (sequence length 512) model and we note that even the Longformer (sequence length 256) model does better than most other PL based models. Interestingly, on the Longformer models, it seems like there was a trend of decreasing CLT performance with increasing sequence length until it bottomed out at the random prediction level.  We expected our PL based models to do way better than they did since there are similarities across programming language structures across languages such as classes, methods, loops, conditionals etc. Our results lead us to think that existing models struggle to learn from the task to generalize to another language. We believe that better predictions on this task might tell us more about the representations learned by LMs.

\section{Conclusion}

\noindent In our work we address the lack of algorithmic complexity datasets by publishing our GFG for C++ and Python. To the extent of our knowledge, this is the first work to include space complexity data. We highlight the importance of dead code elimination and increasing the sequence length to address the issue of tokens that exceed maximum sequence length of LMs. 

\noindent We also introduce a novel CLT task that can be used as an additional benchmark for LM's language comprehension capabilities. Finally, using NMF we demonstrate that LMs can identify large several constructs in the code such as for loop structures, access modifiers and individual functions.

While our indicate results that PL based models fare better on the complexity prediction task, NL based models do better on CLT. However the difference in the performance of NL and PL models is within 5-6\% in both of these cases, so there isn't a clear winner. We'd like to try larger NL and PL models and also larger PL models with a higher max sequence length.

\noindent Based on these results, code complexity prediction using a transformer based approach seems promising. In future work, we plan to extend the GFG dataset by adding more samples for each language and focusing on having a certain minimum number of samples per class. This extension will enable us to perform the fine-tuning and CLT experiments on a larger scale. We would also like to extend the task to be solvable by generative models. 

\section{Ethical Considerations}
All the code on GFG is a under a CCBY license which allows adaptation and remixing as long as we attribute the original authors, while also allowing commercial usage. The code in our dataset contains a comment line identifying the original contributor wherever available.

\bibliographystyle{iclr2023_conference}

\appendix
\section{Appendix}
\subsection{Extant tools to calculate Time and Space Complexities}
\label{extant-complexity-tools}
There are several tools and calculators online that estimate algorithmic time complexity but all of them have been designed with a very specific use-case in mind. Our scraped code typically contains a complete code block with the following components: 
\begin{inparaenum}
  \item A function or class that has the solution to a specific problem, say a function to rotate an array called "Rotate".
  \item Driver code that calls the above class or function and runs the code on all the inputs to be tested.
\end{inparaenum}
None of the tools mentioned below accept the complete code block described above. We do not wish to needlessly aid LMs in any way in their pursuit to understand code, therefore, we expect that our LMs will correctly perform complexity analysis despite these hurdles.

Python libraries such as Radon \citep{radon} and Lizard \citep{Lizard} compute cyclomatic complexity, a quantitative measure of the number
of linearly independent paths in a program’s source code, computed using the control-flow graph of a program. However, the number of linearly independent paths does not
represent how many times these paths were executed.

Big O Notation Calculator \citep{shunnarski3} (Java, Javascript, C, C++, C\#) impressively calculates nested for loop complexities up to a depth of 4 (not tested beyond this depth in this work) but is not designed to be used on on classes or on complete code blocks . 

The python library "big-O-calculator" \citep{pypi-BigO} solely compares time complexities of sorting algorithms.  

A useful python module called big\_O \citep{big_O} empirically computes the time complexity by running a given function over differently sized inputs. We plan to use this module in the future to increase the size of our dataset as it will help us label scraped code which has no associated time complexity. However, this will need additional processing, since it only accepts a Python function as input which necessitates the removal of the driver code from the program. 

Aprove \citep{aprove} was originally designed for termination analysis of
languages like Java, C, Haskell, Prolog. After some re-purposing, it was adapted to also predict time complexities for Java programs. It works best on numeric and heap manipulating programs, successfully classifying time complexity in 73\% of cases (212 total). It is not meant for programs which contain floating point numbers or recursion.

The tools to compute or guess the space complexity of a program are far fewer, the only one we could find was GuessCompx\citep{agenis2019guesscompx}, an R package which performs both time and space analysis empirically. This uses a Generalized Linear Model (GLM) to fit a curve on empirically computed complexity values based on LOO-MSE (leave one out-mean squared error). Furthermore, this approach does not rely on the underlying code at all and uses standard R functions: \texttt{system.time} and \texttt{memory.size} to perform its analysis. Understandably, they do not include NP hard problems in their empirical analysis. We are restricted to analyzing only R functions with this package. We wish to do away with the empirical analysis of time and space altogether, unless it is required to label some missing data as in the case with the big\_O python module.  

The lack of a general tool to solve this problem is not surprising, since we run into Turing's Halting problem. This problem states that given an arbitrary computer program and its input, it is not possible to determine whether the program will finish execution or continue running forever. While deciding whether a simple program will finish execution is straightforward, more complex problems pose a problem. While we can check whether a program halts execution after a few steps, if it does not, it is unknown whether the program will finish execution or continue forever.

Based on the analysis above, we see that none of the tools are directly applicable to the problem that we aim to solve except the big\_O python module which could help us label our missing data provided we only have the functions to solve the problem and exclude the driver code.

\subsection{Dataset construction details}
\label{dataset-construction-details}
We scrape all our data from \href{https://www.geeksforgeeks.org/}{geeksforgeeks (GFG)}. Under the GFG data structures section, we obtain the links to data structure specific problems for arrays, strings, stacks etc. The pages for each of these links contain a list of problems for that specific data structure. We store these these problem links in a dictionary like below:
\begin{lstlisting}[language=python]
dict_problem_urls = {
    "arrays" : [url_problem_a, url_problem_b, ...]
    "trees" : [url_problem_c, url_problem_d, ...] 
}
\end{lstlisting}
We use a combination of beautifulsoup and regular expressions to scrape the code and the associated time \& space complexities from the page. If a page has multiple codes and complexities, we obtain all of them. We ensure that the right code is associated with the right complexities since there could be multiple complexities for a given problem below the code block. To filter code specific to each language, we look for telltale signs associated with a language, such as comments and other HTML based identifiers on the source of a web page, erring on the side of caution. Our scraper give us a pandas dataframe, that contains the URL, code encoded in unicode (to preserve formatting), time and space complexity (as html), the problem type (array / stack / queue) and the comments (success / failure). In this manner we obtain dataframes for C\#, C++, Java, JavaScript and Python. 

To obtain usable code from this dataframe, we decode the unicode data and then perform unicode normalization using the unicodedata library which converts the code along with the requisite formatting into usable text. To obtain the time and space complexities from html, we use regular expressions along with spreadsheet editors, since we needed to look at all our 1410 C++ samples and 1373 Python samples together to check whether a particular regular expression logic worked. There were several cases which could not be tackled even using regular expressions, so we had to manually visit the page and obtain those complexities (approximately 500 samples in each programming language). We were only able to perform this manual step for Python and C++, which is why we only present statistics exclusively for these languages despite also having scraped data in C\#, Java and JavaScript. We also took this opportunity to manually verify the veracity of around 700 samples for Python and C++.

\subsection{Additional statistics about the dataset}
\label{dataset-additional-stats}
\begin{figure}[h]
\begin{center}
\includegraphics[width=1.0\columnwidth]{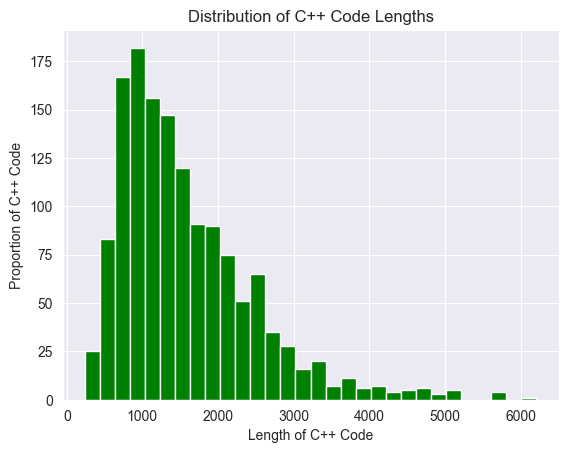}
\end{center}
\caption{Distribution of C++ code lengths, where length is the string based length of code}
\label{cpp-code-length-distribution}
\end{figure}

\begin{figure}[h]
\begin{center}
\includegraphics[width=1.0\columnwidth]{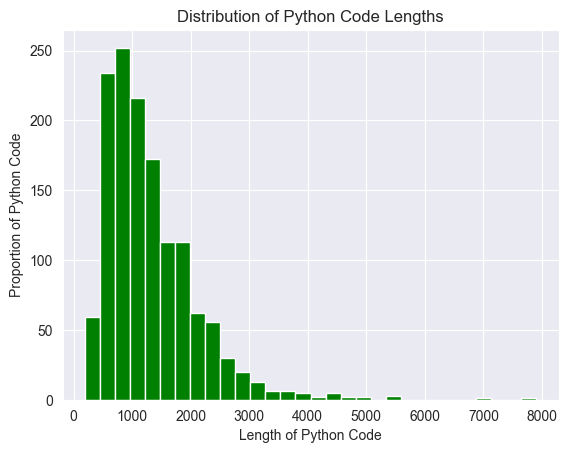}
\end{center}
\caption{Distribution of Python code lengths, where length is the string based length of code}
\label{python-code-length-distribution}
\end{figure}

Our space complexity dataset presented in \hyperref[space-complexity-data-stats]{Table 5} only has 6 classes compared to the 7 classes for time complexity (cubic space complexity was not present in the scraped data). Code length statistics based on length where length is the string length of the code and where length is the number of lines of code are given in \hyperref[code-length-statistics]{Table 6} and \hyperref[code-line-number-statistics]{Table 7} respectively. A visualization of the distribution of code lengths (string length) is given in \hyperref[cpp-code-length-distribution]{Figure 6} and \hyperref[python-code-length-distribution]{Figure 7}.

\begin{table}
\centering
\caption{Space Complexity class statistics}
\label{space-complexity-data-stats}
\begin{tabular}{lcc} 
\toprule
\textbf{Space Complexity} & \multicolumn{1}{l}{\textbf{GFG (C++)}} & \multicolumn{1}{l}{\textbf{GFG (Python)}}  \\ 
\toprule
$O(1)$   & 602    & 603                                        \\
$O(n)$    & 629       & 608                                        \\
$O(n^2)$  & 107         & 96                                         \\
$O(\ln n) $   & 55         & 51                                         \\
$O(n \ln n)$      & 3       & 3                                          \\
NP-hard   & 14            & 12                                         \\ 
\toprule
\textbf{Total}   & 1390       & 1373                                       \\
\bottomrule
\end{tabular}
\end{table}

\begin{table}
\centering
\caption{Code length statistics for the GFG dataset. Here, length is measured akin to string length, as a count of the number of characters in code, not as the number of lines of code.}
\label{code-length-statistics}
\begin{tabular}{ccc} 
\toprule
\textbf{Code length statistic} & \textbf{GFG (C++)} & \textbf{GFG (Python)}  \\ 
\midrule
Mean~                          & 1580.72            & 1323.84                \\
Median~                        & 1365.0             & 1122.0                 \\
Minimum                        & 242                & 195                    \\
Maximum                        & 6203               & 7900                   \\
\bottomrule
\end{tabular}
\end{table}

\begin{table}[h]
\centering
\caption{Code length statistics based on the number of lines of code for the GFG dataset.}
\label{code-line-number-statistics}
\begin{tabular}{ccc} 
\toprule
\textbf{Code length statistic} & \textbf{GFG (C++)} & \textbf{GFG (Python)}  \\ 
\midrule
Mean~                          & 66.84              & 55.05                  \\
Median~                        & 58.0               & 48.0                   \\
Minimum                        & 12                 & 11                     \\
Maximum                        & 223                & 216                    \\
\bottomrule
\end{tabular}
\end{table}

\subsection{Per class accuracies for time and space complexity prediction}
\label{per-class-accuracy}
In \hyperref[per-class-acc-table]{Table 8} , we report the per class test accuracies for our baseline model BERT and our best model GraphCodeBERT on the GFG Python and GFG C++ Datasets, averaged over 5 runs. We observe that our per class accuracies start high and decrease in frequency order for the following classes: linear, quadratic, $n \log n$, $\log n$. Despite there being a similar number of examples (cpp/python) in the $\log n$ (26/25) and constant (33/25), cubic (35/36) and NP-hard classes (28/25), we do not observe a similar level of accuracy across these classes. While it is not surprising that our models have a hard time understanding NP-hard problems, it is surprising that we observe low accuracies on the constant class as well. We hope to remedy this issue in our future work.

\begin{table}
\centering
\caption{Per class accuracies on the GFG C++ and Python datasets. All figures are test accuracy percentages for a particular class, averaged over 5 runs.}
\label{per-class-acc-table}
\begin{tabular}{llccccccc} 
\hline
\textbf{Model} & \textbf{Dataset} & \textbf{$O(1)$} & \textbf{$O(n)$} & \textbf{$O(logn)$} & \textbf{$O(n^2)$} & \textbf{$O(n^3)$} & \textbf{$O(nlogn$)} & \textbf{NP}  \\ 
\hline
BERT           & GFG - C++        & 5.7             & 86.8            & 40.0               & 76.5              & 14.2              & 64.0                & 0            \\
GraphCodeBERT  & GFG - C++        & 0               & 89.5            & 36.0               & 80.0              & 20.0              & 68.0                & 0            \\ 
\hline
BERT           & GFG - Python     & 8.5             & 87.7            & 28.0               & 75.8              & 5.71              & 68.3                & 0            \\
GraphCodeBERT  & GFG - Python     & 8.5             & 91.0            & 28.0               & 75.8              & 17.1              & 66.6                & 0            \\
\hline
\end{tabular}
\end{table}

\end{document}